\documentclass[prd,preprintnumbers,showkeys,nofootinbib]{article}
\usepackage{bm, graphicx,amsmath,amssymb,epsf}

\begin{document}


\title{\Large \bf  Duality symmetry in high energy scattering }
\author{\large A.~Prygarin
\bigskip \\
{\it  II. Institute for Theoretical Physics, Hamburg University, Germany} \\
 }

\maketitle

\begin{abstract}
We discuss the duality  symmetry of  the linear~(BFKL) and the non-linear~(BK) high energy evolutions in the multicolor limit. We show that the usual color dipole picture is dual to the forward reggeized gluon formulation. The presented analysis is also generalized to the non-forward case where we suggest 
an extended version of the  duality symmetry. We give it a physical interpretation as a symmetry under rotation of the Kernel in the transverse space from $s$-channel~(dipoles) to $t$-channel~(reggeized gluons). The duality symmetry is related to the integrability of the system. The duality symmetry of the BK equation found in the present study can be regarded as an indirect indication of its integrability.

\end{abstract}

\section{Introduction}
\label{sec:intro}
\hspace{0.4cm} The high energy scattering is characterized  by Regge kinematics where the transferred momentum is much smaller than the center-of-mass energy of the scattered particles. The analysis of particle scattering in the Regge limit $|t/s|\ll 1$ led to the introduction of Reggeons exchanges in the $t$-channel which correspond to Regge trajectories $\alpha(t)\simeq \alpha_0+\alpha' t$ in power energy behavior of the scattering amplitude $A \sim (s/s_0)^{\alpha(t)}$. The Reggeon with intercept $\alpha_0$ close to zero, called Pomeron, is leading at high energies.

In QCD the Pomeron is described by Balitsky-Fadin-Kuraev-Lipatov~(BFKL) \cite{BFKL} evolution equation for a composite state of two reggeized gluons in $t$-channel. The reggeization of gluons modifies the gluon propagator by $(s/s_0)^{\alpha(t)}$ and is a consequence of the "bootstrap" condition which implies that a system two or more reggeized gluons projected on color octet state collapses to one reggeized gluon. The BFKL evolution equation resums contributions of $\alpha_s \log s$ where the QCD coupling constant $\alpha_s$ is accompanied by the same power of the logarithm of energy~(Leading Logarithm Approximation-LLA). This resummation assumes multi-Regge kinematics of $s$-channel gluons with their large relative separation in rapidity.

The BFKL equation was solved by Lipatov \cite{BFKLsol} exploiting the two dimensional conformal invariance of the BFKL Kernel. It was also noticed \cite{Lipatov:1998as} that the conformal BFKL Hamiltonian has an additional symmetry, namely, it is invariant under a change of variables $k_i=x_i-x_j$, where $k$ is the transverse momentum of the $i$th reggeized gluon and $x$ is the Fourier conjugate to it coordinate. This symmetry, named \emph{the duality symmetry}, is a result of the integrability of the BFKL equation.

 The dipole model is an alternative description of the BFKL evolution in the coordinate space using $s$-channel unitarity developed by Mueller \cite{MUCD}. In the limit of a large number of colors $N_c$ a gluon can be viewed as a double quark-antiquark line and only the nearest neighbor interactions survive. The dipole model was formulated for a color singlet scattering amplitude in light-cone gauge, while the momentum BFKL is valid for any color state and an arbitrary number of colors $N_c$. Despite the fact that the dipole formulation appears as a special case of the BFKL evolution, it has a  very intuitive physical picture of the underlying evolution in terms of dipole splittings. The full correspondence between the two formulations is still not fully clear, in particular, the exact relation between gluon reggeization and dipole wave function renormalization was not completely verified. Both, gluon reggeization and dipole renormalization come from virtual contributions, but by virtue of the Fourier transform the virtual contributions are \emph{partly} transformed into the real emission terms.

In the present study we address the duality symmetry, we show that  it can be viewed as a duality between the reggeized gluon and dipole formulations of the linear BFKL evolution. 
We find that the duality symmetry also holds for the non-linear BK evolution equation provided we impose a condition dual to the dipole size conservation.  The clear physical meaning of duality symmetry as a symmetry between reggeized gluon formulation and  the dipole picture provides a corresponding UV/IR duality, which is  likely to hold at NLO BFKL in $\mathcal{N}=4$ super Yang-Mills and beyond.

The two dimensional duality symmetry of the BFKL and the BK equations can further be related to the four dimensional dual conformal symmetry used in Wilson Loops/Scattering Amplitude duality~\cite{Drummond:2007aua} as well as to $T$-duality exploited in AdS/CFT calculations of gluon scattering amplitudes~\cite{Alday}.

The paper organized as follows. 

At the beginning we present a brief introduction to the BFKL equation in the conformal basis and explain the meaning of the duality symmetry in this formalism. 
Next, we demonstrate the duality symmetry as a symmetry between the BFKL equation in the momentum space and its dipole formulation. In the upcoming section we show that it extends to the non-linear case of the BK equation. Then we give a physical interpretation of the duality symmetry for the non-forward evolution. Some calculations are presented in the Appendix.

\section{BFKL in the conformal basis and the duality symmetry}
\label{sec:BFKLconformal}
  \hspace{0.4cm} In the scattering at very high energies  the scattering amplitude in the Leading Logarithmic Approximation~(LLA) is obtained by summing contributions of the order of $(g^2 \ln(s))^n$, where $g$ is the coupling constant. In this limit $t$-channel gluons reggeize and the BFKL pomeron appears as a composite state of two reggeized gluons.

The relevant kinematics is called multi-Regge kinematics and it is characterized by the factorization of the longitudinal and transverse degrees of freedom. Due to this factorization the BFKL pomeron can be written as a state in the two dimensional transverse space that evolves with rapidity which plays a role of an imaginary time. This fact makes it possible to formulate the color singlet BFKL dynamics in form of the Schr\"odinger equation for the wave function $f_{m,\tilde{m}}(\vec{\rho}_1,\vec{\rho}_2,...,\vec{\rho}_n;\vec{\rho}_0)$ for a system of $n$-reggeized gluons \cite{Kwiecinski:1980wb,Bartels:1980pe,Lipatov:1990zb}, the BFKL equation is obtained for $n=2$.
The vectors $\vec{\rho}_k$ are two dimensional coordinates of the reggeized gluons, and $m$ and $\tilde{m}$ are  the conformal weights 
\begin{eqnarray}
 m=\frac{1}{2}+i\nu+\frac{n}{2} ,\;\;\; \tilde{m}=\frac{1}{2}+i\nu-\frac{n}{2}
\end{eqnarray}
 which are expressed in terms of the anomalous dimension $\gamma=1+2i\nu$ and the integer conformal spin $n$. The anomalous dimension and the conformal spin in this context were introduced when solving the BFKL equation in the complex coordinates 
\begin{eqnarray}
 \rho_k=x_k+iy_k, \;\;\; \rho^*_k=x_k-iy_k 
\end{eqnarray}
using the conformal properties of the BFKL Kernel. 

The BFKL wave function $f_{m,\tilde{m}}$ satisfies the Schr\"odinger equation
\begin{eqnarray}
 E_{m,\tilde{m}}f_{m,\tilde{m}}=H f_{m,\tilde{m}}
\end{eqnarray}
with the energy $E_{m,\tilde{m}}$ being proportional to the position of the singularity in the complex angular momentum $j$ plane. In the multicolor limit the Hamiltonian possesses a property of holomorphic separability 
\begin{eqnarray}
 H=\frac{1}{2}\left(h+h^*\right)
\end{eqnarray}
where the holomorphic and the anti-holomorphic Hamiltonians 
\begin{eqnarray}\label{hamsmall}
 h=\sum_{k=1}^n h_{k,k+1}, \;\;\; h^*=\sum_{k=1}^n h^*_{k,k+1}
\end{eqnarray}

are expressed through the BFKL operator \cite{Lipatov:1993qn}
\begin{eqnarray}\label{hholom}
 h_{k,k+1}=\log(p_k)+\log(p_{k+1})+\frac{1}{p_k}\log(\rho_{k+1})p_k+\frac{1}{p_{k+1}}\log(\rho_{k+1})p_{k+1}+2\gamma
\end{eqnarray}
In Eq.~(\ref{hholom}) one defines $\rho_{k,k+1}=\rho_k-\rho_{k+1}$, $p_k=i\partial/(\partial \rho_k)$, $p^*_k=i\partial/(\partial \rho^*_k)$ and $\gamma=-\psi(1)$~(the Euler constant). The holomorphic separability of the Hamiltonian means the holomorphic factorization  of the wave function 
\begin{eqnarray}\label{confbfkl}
f_{m,\tilde{m}}(\vec{\rho}_1,\vec{\rho}_2,...,\vec{\rho}_n;\vec{\rho}_0)=\sum_{r,l} c_{r,l}f^r_{m}(\rho_1,\rho_2,...,\rho_n;\rho_0)
f^l_{\tilde{m}}(\rho^*_1,\rho^*_2,...,\rho^*_n;\rho^*_0)
\end{eqnarray}
 and the Schr\"odinger equations in the holomorphic and the anti-holomorphic spaces 
\begin{eqnarray}
 \epsilon_m f_m=h f_m, \;\;\; \epsilon_{\tilde{m}}f_{\tilde{m}}=h^* f_{\tilde{m}}, \;\;\;E_{m,\tilde{m}}=\epsilon_m+\epsilon_{\tilde{m}}
\end{eqnarray}
The degenerate solutions are accounted for by the coefficients $c_{r,l}$ in Eq.~(\ref{confbfkl}), which are fixed by the singlevaluedness condition for the wave function in the two dimensional space.

It is interesting to note that the BFKL way function can be normalized in two different ways
\begin{eqnarray}
 \parallel f\parallel^2_1=\int \prod_{r=1}^n d^2 \rho_r \left| \prod^n_{r=1} \rho^{-1}_{r,r+1} f\right|^2,
\;\;\;
 \parallel f\parallel^2_2=\int \prod_{r=1}^n d^2 \rho_r \left| \prod^n_{r=1} p_{r} f\right|^2
\end{eqnarray}
This is in an agreement with the hermicity properties of the Hamiltonian, since the transposed Hamiltonian $h^t$ can be obtained by two different similarity transformations   \cite{Lipatov:1993yb}
\begin{eqnarray}\label{norm}
 h^t=\prod_{r=1}^n p_r h \prod_{r=1}^n p_r^{-1}=\prod_{r=1}^n \rho_{r,r+1}^{-1} h \prod_{r=1}^n \rho_{r,r+1}
\end{eqnarray}
 The Hamiltonian should be symmetrical under change of variables 
\begin{eqnarray}\label{change}
 \rho_{k-1,k} \to p_{k} \to \rho_{k,k+1}
\end{eqnarray}
accompanied by the change of the operator ordering. Indeed, this property becomes obvious if we rewrite the Hamiltonian Eq.~(\ref{hamsmall}) in the form of 
\begin{eqnarray}
 h=h_p+h_{\rho}
\end{eqnarray}
with
\begin{eqnarray}\label{hp}
h_p=\sum_{k=1}^n \left( \log(p_k) +\frac{1}{2} \sum_{\lambda=\pm1}  \rho_{k,k+\lambda} \log(p_k) \rho^{-1}_{k,k+\lambda}+\gamma \right)
\end{eqnarray}

and 
\begin{eqnarray}\label{hrho}
h_\rho=\sum_{k=1}^n \left( \log(\rho_{k,k+1}) +\frac{1}{2} \sum_{\lambda=\pm1}  p^{-1}_{k+(1+\lambda)/2} \log(\rho_{k,k+1}) p_{k+(1+\lambda)/2}+\gamma \right)
\end{eqnarray}

The invariance of the BFKL Hamiltonian under the change of the variables Eq.~(\ref{change}) together with  the change of the operator ordering was called by Lipatov the \emph{duality} symmetry. 
The \emph{duality} symmetry   implies that the BFKL Hamiltonian commutes $[h,A]=0$ with the differential operator
\begin{eqnarray}
 A=\rho_{12}\rho_{23}...\rho_{n1}p_1p_2...p_n.
\end{eqnarray}
 or, more generally, there is a family of mutually commuting integrals of motion \cite{Lipatov:1993yb}
\begin{eqnarray}
 [q_r,q_s]=0, \;\;\; [q_r,h]=0
\end{eqnarray}
and they are given by 
\begin{eqnarray}
 q_r=\sum_{i_1<i_2<...<i_r} \rho_{i_1,i_2}\rho_{i_2,i_3}...\rho_{i_r,i_1}p_{i_1}p_{i_2}... p_{i_r}.
\end{eqnarray}
The operators $q_r$  build a complete set of the invariants of the transformation. Therefore the Hamiltonian $h$ is their function
\begin{eqnarray}
 h=h(q_2,q_3,...,q_n)
\end{eqnarray}
and a common eigenfunction of $q_r$ is simultaneously a solution to the Schr\"odinger equation. This fact explains why the duality symmetry is related to the integrability of the a system of Reggeons in the limit of the large number of colors $N_c$. In the the  multicolor limit only nearest neighbor interactions are not suppressed and  the BFKL dynamics is similar to that of Ising spin chain model.

The transformation Eq.~(\ref{change}) of the holomorphic BFKL Hamiltonian is an unitary transformation only for a vanishing total momentum
\begin{eqnarray}
 \vec{p}=\sum^n_{r=1}\vec{p}_r
\end{eqnarray}
which guarantees the cyclicity of the momenta $p_r$ important for their representation by the difference of coordinates $\rho_{r,r+1}$. For the composite state of two reggeized gluons~(usual BFKL case) for $n=2$, this can be achieved only for the zero transferred momentum $\vec{q}=0$. Only in this case one can really identify the dual coordinates  $\vec{\rho}_{r,r+1}$ of the momenta $\vec{p}_r$ with their conjugate coordinates. In a more general case these two is not the same object. However, the integrability of the non-forward BFKL suggests that the duality symmetry should be present also in the case of $\vec{q}\neq 0$, but in an implicit way. The possible extension of the duality symmetry of the BFKL for $\vec{q}\neq 0$ is discussed in section \ref{sec:nonforw}.

The duality symmetry appears as a special case of a more general \emph{dual conformal} symmetry, which is a usual conformal symmetry of a system in the dual space, where the momenta are parametrized by the \emph{dual} coordinates $p_r=x_{r}-x_{r+1}$. The \emph{dual} coordinates $x_r$ have dimensions of mass and \emph{a priori} are not related to the conjugate coordinates $\rho_{r}$ of the corresponding momenta.  The dual conformal symmetry~(in four dimensions) plays the central role in recent developments in calculating multiloop amplitudes with many legs using Wilson Loop/Scattering Amplitude duality \cite{Drummond:2007aua}. It is commonly believed now that this more general \emph{dual conformal} symmetry is responsible for the integrability of the amplitudes in (super)conformal $\mathcal{N}=4$ SYM gauge theory in the multicolor limit.  It is worth emphasizing that in the leading order BFKL the multiregge kinematics selects only gluons, the contributions of other particles are suppressed due the powerlike behavior of the amplitude $s^{j-1}$, where $j$ is the angular momentum of the exchanged particle.  This means that in the multiregge kinematics QCD is not much different from $\mathcal{N}=4$ SYM, and the factorization of the longitudinal~(rapidity) and the transverse momenta explains why the duality symmetry of the BFKL dynamics in two dimensions should be related to the four-dimensional dual conformal symmetry  in multicolor $\mathcal{N}=4$ SYM amplitudes.

From the strong coupling side, the gluon scattering amplitudes were calculated using gauge/string duality or AdS/CFT correspondence and the crucial assumption was the validity of the fermionic $T$-duality. Applying fermionic $T$-duality in the form very similar to the dual coordinates the authors calculated the  classical string solution for a four point scattering amplitude, which agrees with the Bern-Dixon-Smirnov~(BDS) ansatz \cite{Bern:2005iz}, WL amplitudes and the solution to the BFKL equation. Further analysis of amplitudes with a higher number of external legs showed the validity of this assumption. This means that the fermionic $T$-duality from a large coupling side and the dual conformal symmetry from the weak coupling constant perturbation theory are related to each other.

Both the \emph{duality} symmetry and the \emph{dual conformal } symmetry appear like a mathematical peculiarity lacking any physical interpretation. 
The main objective of the present study is to give a physical interpretation of the duality symmetry of the BFKL approach as a symmetry between the reggeized gluon formulation and the dipole picture. In particular, we show that the duality symmetry can be viewed as a symmetry under rotation of the BFKL Kernel in the transverse plane from $s$-channel to $t$-channel and back. This suggests a similar interpretation for the dual conformal symmetry of the scattering amplitudes in an arbitrary kinematics. In the next section we consider the forward BFKL equation and show its selfduality with a proper choice of the Fourier transform. This analysis is then extended to the non-linear BK equation.

\section{Forward BFKL}
\label{sec:BFKL}

\begin{figure}[h]
\begin{center}
\includegraphics[height=1in]{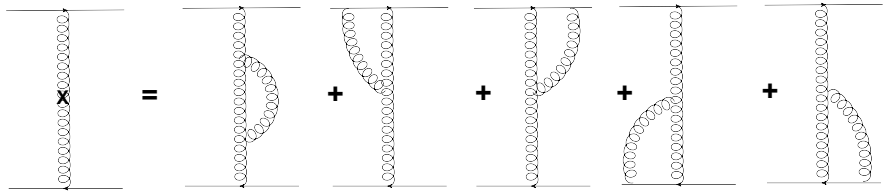}  
\end{center}
\caption{The virtual part of the BFKL kernel appears due to virtual corrections to the propagator of $t$-channel gluon. In the Regge limit the gluon reggeize and its trajectory effectively includes all virtual Feyman diagrams. The small cross on the gluon denotes that fact that the gluon is reggeized.} \label{fig:BFKLvirtual}
\end{figure}

 \hspace{0.4cm} In this section we consider a special case of the BFKL equation of the zero transferred momentum $q=0$\footnote{From now on we deal only with two dimensional transverse momenta and  coordinates in Euclidean space~(~$k^2=\vec{k}^2_\perp$~and~$x^2=\vec{x}^2_\perp$~)  and do not introduce any special notation for them.}. We show that it is dual to the dipole linear evolution equation for a fixed impact parameter $b$. It is worth emphasizing that a usual Fourier transform implies integration over the impact parameter for $q=0$, and despite the fact that it makes no difference in our analysis for the linear case, it will be shown later that the right way to identify the duality it is to keep the impact parameter $b$ fixed. The general form of the multicolor, singlet BFKL equation reads
\begin{eqnarray}\label{BFKL}
\left(\frac{\partial}{\partial y}-\epsilon(-k^2)-\epsilon(-(k-q)^2)\right)\mathcal{F}(k,q)
=\frac{\alpha_s N_c}{2\pi^2} \int d^2 \chi \frac{K(k,\chi)}{\chi^2(\chi-q)^2}\mathcal{F}(\chi,q)
\end{eqnarray}
where $\epsilon(-k^2)$  the gluon Regge trajectory given by
\begin{eqnarray}\label{regge}
\epsilon(-k^2)=-\frac{\alpha_s N_c}{4\pi^2}\int d^2 \chi \frac{k^2}{\chi^2(\chi-k)^2}
\end{eqnarray}
accounts for virtual contributions coming from the  loop corrections to the propagator of $t$-channel gluons illustrated in Fig.~\ref{fig:BFKLvirtual}.

\begin{figure}[h]
\begin{center}
\includegraphics[height=1in]{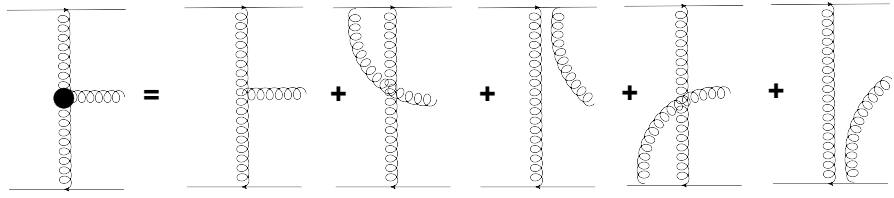}  
\end{center}
\caption{The effective production vertex that builds the real part of the BFKL kernel, consists of several Feyman diagrams. In the Regge limit the real gluon emission factorizes and is independent on the properties of the scattered particles. The dark blob denotes the effective Lipatov vertex.} \label{fig:lipatovVertex}
\end{figure}

 The amplitude $\mathcal{F}(k,q)$ is  always a function of rapidity $y$ though it is not reflected in our notation.
The function $K(k,\chi)$ is the part of the BFKL Kernel, which describes real gluon emissions and is obtained from the  square the effective Lipatov vertex. It is given by
\begin{eqnarray}\label{realKernel}
K(k,\chi)=q^2-\frac{k^2(\chi-q)^2}{(\chi-k)^2}-\frac{\chi^2(k-q)^2}{(\chi-k)^2}.
\end{eqnarray}
The effective Lipatov vertex is a gauge invariant object that includes several Feyman diagrams shown in Fig.~\ref{fig:lipatovVertex}.

\begin{figure}[h]
\begin{center}
\includegraphics[height=1.5in]{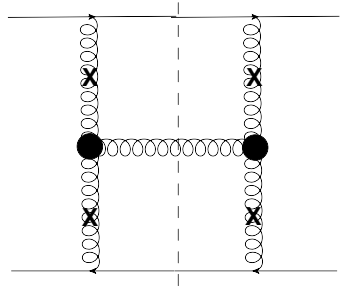}  
\end{center}
\caption{The figure illustrates the full BFKL Kernel. The real part comes from effective production vertices denoted by dark blobs, while the virtual contribution are accounted for by the reggeization of $t$-channel gluons. The crosses on $t$-channel gluons reflect the fact that they are reggeized.} \label{fig:BFKL}
\end{figure}

The full BFKL Kernel consisting of real and virtual contributions is thus obtained by taking a square of the real part and including the virtual contributions for $t$-channel gluons as shown in Fig.~\ref{fig:BFKL}.

Using these definitions in Eq.~(\ref{BFKL}) one can write the singlet BFKL equation as

\begin{eqnarray}\label{BFKLfull}
\frac{\partial \mathcal{F}(k,q)}{\partial y}
=\frac{\alpha_s N_c}{2\pi^2} \int d^2 \chi \left(\frac{k^2}{\chi^2(\chi-k)^2}+\frac{(k-q)^2}{(\chi-q)^2(\chi-k)^2}-\frac{q^2}{\chi^2(\chi-q)^2}\right)\mathcal{F}(\chi,q)
\\
- \frac{\alpha_s N_c}{4\pi^2} \int d^2 \chi \frac{k^2}{\chi^2(\chi-k)^2}\mathcal{F}(k,q)
- \frac{\alpha_s N_c}{4\pi^2} \int d^2 \chi \frac{(k-q)^2}{\chi^2(\chi-k+q)^2}\mathcal{F}(k,q) \nonumber
\end{eqnarray}
the first three terms in brackets on r.h.s. describe the real gluon emissions, while the last two terms come from virtual emissions and are responsible for the reggeization of gluon. In the forward case~($q=0$) Eq.~(\ref{BFKLfull}) reduces to

\begin{eqnarray}\label{BFKLforward}
\frac{\partial \mathcal{F}(k)}{\partial y}
=\frac{\alpha_s N_c}{2\pi^2} \int d^2 \chi \left(2\frac{k^2}{\chi^2(\chi-k)^2}\right)\mathcal{F}(\chi)
- 2\frac{\alpha_s N_c}{4\pi^2} \int d^2 \chi \frac{k^2}{\chi^2(\chi-k)^2}\mathcal{F}(k).
\end{eqnarray}

In the dipole model formulation the linear evolution equation reads
\begin{eqnarray}\label{BFKLdipole}
\frac{\partial N(x_{12},b_{12})}{\partial y}
=\frac{\alpha_s N_c}{2\pi^2} \int d^2 x_{13} \frac{x_{12}^2}{x_{13}^2 x_{32}^2}
\left\{ N(x_{13},b_{13})+N(x_{32},b_{32})-N(x_{12},b_{12})\right\}
\end{eqnarray}
where  $N(x_{12},b_{12})$ is the imaginary part of the scattering amplitude of a colorless dipole with transverse coordinates $x_1$ and $x_2$ of its quark and antiquark components. For simplicity, we do not write explicitly the rapidity argument in the energy dependent function $N(x_{12},b_{12})$.  The transverse size and the impact parameter  of the dipole are $x_{12}=x_1-x_2$ and $b_{12}=(x_1+x_2)/2$ respectively. The first two terms of r.h.s are responsible for the real "early" emissions, where the dipole splitting occurs before the interaction with the target in the light-cone time. These are to be confronted with the "late" real emissions of a dipole being split after the interaction happens, which cancel out with corresponding virtual contributions. For a detailed discussion of diagrammatic and functional meaning of real-virtual cancellation one is referred to \cite{Levin:1900tt}. The last term of Eq.~(\ref{BFKLdipole}) comes from the "early" virtual emissions, where the splitting and the merging of the dipole happens before the interaction with the target. The role of this contribution is renormalize the wave function of the initial dipole.

In the remaining part of the present section and the next section we ignore the dependence on the impact parameter, which is equivalent to integrating it out or to simply setting $b=0$. Both of the cases lead to the same expression for the linear BFKL equation provided one changes properly the normalization of the scattering amplitude, while for  the non-linear BK equation they are obviously very much different. For our purposes we set $b=0$ here and thoroughly discuss  the dependence on the impact parameter in section \ref{sec:nonforw}.   

It is easy to see to that ignoring  $b$-dependence  in Eq.~(\ref{BFKLdipole}) and noting that the first and the second terms of r.h.s can be brought to the same form by redefinition of  the integration variable, the equation in Eq.~(\ref{BFKLdipole}) reads

\begin{eqnarray}\label{BFKLdipolezero}
\frac{\partial N(x_{12})}{\partial y}
=\frac{\alpha_s N_c}{2\pi^2} \int d^2 x_{13} \frac{x_{12}^2}{x_{13}^2 x_{32}^2}
\left\{ 2N(x_{13})-N(x_{12})\right\}.
\end{eqnarray}
For the sake of simplicity we do not introduce a new notation for the scattering amplitude and define $N(x,b=0)\equiv N(x)$. The immediate analogy between Eq.~(\ref{BFKLforward}) and Eq.~(\ref{BFKLdipolezero}) suggest the duality symmetry
\begin{eqnarray}\label{duality}
k \Leftrightarrow x_{12}
\end{eqnarray}
 that takes reggeized gluon formulation to the dipole picture and back. The duality means that the direct substitution of the initial dipole size $x_{12}$ for  the upper virtuality of the reggeized gluons $k$ building the BFKL ladder coincides with the form of Fourier transformed BFKL, provided  $k$ and $x_{12}$ are Fourier conjugate. A commonly used form of the Fourier transform is defined as
 \begin{eqnarray}\label{Fourier}
 \mathcal{F}(k)\equiv \frac{1}{(2\pi)^2} \int \frac{d^2 x_{12}}{x^2_{12}} \;e^{-i k \; x_{12}} N(x_{12}) 
 \end{eqnarray}
keeps the Fourier transformed amplitude dimensionless, preserves conformal invariance, but hides the duality symmetry.  For our purposes, an appropriate way to do this is to take advantage of the conformal invariance of the scattering amplitude and to define dimensionless coordinates as follows
\begin{eqnarray} \label{xcoord}
 \rho_{ij}=\frac{x_{ij}}{|x_{12}|} 
\end{eqnarray}
and 
\begin{eqnarray}\label{kcoord}
  \chi_i=\frac{k_i}{|k|} \;\;\; \text{with} \;\;\;  \kappa= \frac{k}{|k|}.
\end{eqnarray}

The Fourier transform in these coordinates reads \footnote{we change notation $\mathcal{F} \to \mathcal{N}$ needed for the non-linear case to make the presentation clear}

 \begin{eqnarray}\label{confFourier}
 \mathcal{N}(\kappa)\equiv \frac{1}{(2\pi)^2} \int d^2 \rho_{12} \;e^{-i \kappa \; \rho_{12}} N(\rho_{12}) .
 \end{eqnarray}

With the new definition of the Fourier transform the dipole evolution with omitted impact parameter dependence Eq.~(\ref{BFKLdipolezero}) transforms to the forward BFKL given by Eq.~(\ref{BFKLforward}). It is a well known fact that such a Fourier transform and its inverse bring in contributions proportional to the $\delta$-functions either of the initial dipole size $\delta^2(x_{12})$ or of the upper virtuality of the BFKL ladder $\delta^2(k)$. It was shown by Mueller and Tang \cite{Mueller:1992pe} that they are safely removed due to vanishing impact factors when $k$ or $x_{12}$ go to zero. For a  complete discussion of the Fourier transform of the BFKL equation see Bartels et al. \cite{Bartels:1995rn} and Forshaw $\&$  Ryskin \cite{Forshaw:1995ax}.

In the next section we consider the BK evolution equation, which includes a non-linear term that corresponds to the triple Pomeron vertex. We show that the BK equation also enjoys the dual  symmetry provided one imposes a constraint in the momentum space that is analogous to the fact that dipole size does not change during the interaction, which is the basic assumption of the dipole model.

\section{Non-linear evolution}
\label{sec:nonlin}
\hspace{0.4cm} The non-linear evolution was formulated by Balitsky \cite{B} in Wilson Lines formalism, and soon after that independently by Kovchegov \cite{K} in the dipole picture.  In the multicolor limit they coincide provided one neglects high order correlations in multiple rescattering of $q \bar{q}$ pair. The Balitsky-Kovchegov equation reads

\begin{eqnarray}\label{BK}
\hspace{-0.5cm}
\frac{\partial N(x_{12},b_{12})}{\partial y}
=\frac{\alpha_s N_c}{2\pi^2} \int d^2 x_{13} \frac{x_{12}^2}{x_{13}^2 x_{32}^2}
\left\{ N(x_{13},b_{13})+N(x_{32},b_{32})-N(x_{12},b_{12})-N(x_{13},b_{13})N(x_{32},b_{32})\right\}.
\end{eqnarray}

The linear terms on r.h.s of Eq.~(\ref{BK}) reproduce the BFKL equation for small values of the scattering amplitude and are responsible for the BFKL Pomeron propagation. 
The non-linear term describes the situation where two newly formed dipoles interact independently with the target and describe unitarization corrections corresponding to the triple Pomeron vertex.

A few words to be said about the initial condition of the BK equation. It has an eikonal form of
\begin{eqnarray}\label{initBK}
 N(x)=1-e^{\frac{-\sigma(x)}{2}}
\end{eqnarray}
where $\sigma(x)$ is the total cross section of a scattering of a dipole with transverse size $x$. The initial condition Eq.~(\ref{initBK}) is not conformal invariant and thus cannot be invariant in a dual way, in contrast to the initial condition of the BFKL equation. However, we are interested in the pure evolution for an arbitrary initial condition, regardless its properties. In fact, this is not completely true, since, by construction, the BK equation account for the "bootstrap" property of the initial condition, which means that quadratic terms of the same argument $N^2(x)$ are absorbed into corresponding linear terms and never appear in the evolution equation. 

To show the duality symmetry of the BK equation, we have only two extra pieces in addition to those of the BFKL to be Fourier transformed. Namely, the  squared dipole splitting term 
\begin{eqnarray}\label{squaresplit}
 \int d^2 x_{13}\frac{1}{x^2_{13}}N(x_{13})N(x_{32})
\end{eqnarray}
and the crossed splitting term

\begin{eqnarray}\label{crossedsplit}
 \int d^2 x_{13}\frac{x_{13}}{x^2_{13}}\frac{x_{23}}{x^2_{23}}N(x_{13})N(x_{32}).
\end{eqnarray}
Other two terms are obtained by virtue of $1 \leftrightarrow 2$ symmetry of the BK equation.  In the Appendix we present a detailed calculation  of the Fourier transform and here we only want to emphasize some important points.  The traditional Fourier transform Eq.~(\ref{Fourier}) takes the non-linear term of Eq.~(\ref{BK}) to a full square of some function $\mathcal{\phi}(k)$ of the transverse momentum of the initial $q\bar{q}$ state. One immediately faces a problem in giving this squared term $\mathcal{\phi}^2(k)$ a physical interpretation in terms of reggeized gluons, since the amplitude corresponding to a colorless state of reggeized gluons obeys the "bootstrap" condition similar to the BK amplitude.
Such a quadratic term should never appear in the evolution equation written for Pomeron fan diagrams in momentum space for an amplitude with bootstrap. In the Gribov-Levin-Ryskin~(GLR) equation \cite{GLR} the non-liner term $\mathcal{\phi}^2(k)$ was introduced because its initial condition corresponds to one Pomeron exchange. To obtain the non-linear evolution equation for that initial condition, one should consider a non-truncated version of the BK equation with a double  interaction of the same dipole. This problem is beyond the scope of the present paper and its detailed study will be published by us elsewhere.

To demonstrate how this problem is solved in our case we consider the crossed splitting term Eq.~(\ref{crossedsplit}). In Appendix we find its  Fourier transform in dimensionless coordinates Eq.~(\ref{xcoord}) and Eq.~(\ref{kcoord}) 
\begin{eqnarray}\label{crossedFourier}
  \int d^2 \chi_1 d^2 \chi_2 \frac{\chi_1}{\chi^2_1}\frac{\chi_2}{\chi^2_2}\mathcal{N}(\chi_1-\kappa)\mathcal{N}(\chi_2+\kappa).
\end{eqnarray}
This terms is usually interpreted as a squared term. One can see that it contains one extra integration compared to Eq.~(\ref{crossedsplit}). The basic assumption of the dipole model is that the dipole size is preserved during multiple rescatterings. In particular, this means that $x_{13}-x_{23}=x_{12}$ ($\rho_{13}-\rho_{23}=\rho_{12}$ ), which is to be translated into $\chi_1-\chi_2=\kappa$ where $\chi_1$ and $\chi_1$ are conjugate to $x_{13}$ and $x_{23}$, respectively. Using this condition we can integrate over $\chi_2$  

\begin{eqnarray}\label{crossedFourierFin}
  \int d^2 \chi_1  \frac{\chi_1}{\chi^2_1}\frac{\chi_1-\kappa}{(\chi_1-\kappa)^2}\mathcal{N}(\chi_1-\kappa)\mathcal{N}(\chi_1).
\end{eqnarray}
Identifying $\chi_1\rightarrow \rho_{13}$, $\kappa\rightarrow \rho_{12}$ and $\mathcal{N}(\chi) \rightarrow N(\rho_{12})$ this reads 

\begin{eqnarray}\label{crossedFourierFin}
  \int d^2 \rho_{13}  \frac{\rho_{13}}{\rho^2_{13}}\frac{\rho_{23}}{\rho^2_{23}}N(\rho_{13})N(\rho_{32})=\int d^2 x_{13}  \frac{x_{13}}{x^2_{13}}\frac{x_{23}}{x^2_{23}}N(x_{13})N(x_{32})
\end{eqnarray}

due to conformal invariance of the scattering amplitude.

The crossed term is unique in the sense that it is dual to itself, while 
the Fourier  transform of the squared splitting term Eq.~(\ref{squaresplit}) 
vanishes provided 
one imposes the condition dual to the dipole size conservation mentioned 
before. Other linear 
terms are mixed when Fourier transformed, but their sum enjoys self-duality. 
This way we show that the BK equation has the duality symmetry, obvious for the multicolor BFKL equation. This is one of the main results of this paper. The equation dual to the BK equation Eq.~(\ref{BK}) with a fixed~(set zero) impact parameter reads

\begin{eqnarray}\label{dualBK}
\hspace{-0.5cm}
\frac{\partial \mathcal{N}(\kappa)}{\partial y}
=\frac{\alpha_s N_c}{2\pi^2} \int d^2 \chi \frac{\kappa^2}{\chi^2 (\chi-\kappa)^2}
\left\{ \mathcal{N}(\chi)+ \mathcal{N}(\chi-\kappa)- \mathcal{N}(\kappa)- \mathcal{N}(\chi) \mathcal{N}(\chi-\kappa)\right\}
\end{eqnarray}
with the $ \mathcal{N}(\kappa)$ given by Eq.~(\ref{confFourier}).

The duality symmetry is related to integrability and imposes an additional constraint on the eigenfunctions. This suggests that the BK equation is fully integrable system for a fixed impact parameter. 
The selfduality of the BK equation shown in the present study seems to be related to an interesting property of the triple pomeron vertex found  by Korchemsky in \cite{Korchemsky:1997fy}. This property was called by the author  \emph{a crossing symmetry} of the $2\to 4$ reggeized gluon vertex in the conformal basis for large $N_c$. This sort of the crossing symmetry gives rise to an operator algebra, which suggests to interpret the triple pomeron vertex as a three-point function in some two dimensional conformal field theory. The analysis of Ref.~\cite{Korchemsky:1997fy} echoes our conclusion that the BK equation should be integrable. This point certainly requires further investigation and will be addressed by us in the future studies.

In the next section we discuss in more details the physical meaning of the duality symmetry and show how it relates the dipole picture and the reggeized gluon description of the high energy evolution.  We also consider the non-forward case of the BFKL dynamics and suggest the extension of a single variable duality symmetry considered above.

\section{Non-forward case }
\label{sec:nonforw}

\begin{figure}[h]
\begin{center}$
\begin{array}{ccc}
\includegraphics[width=2in]{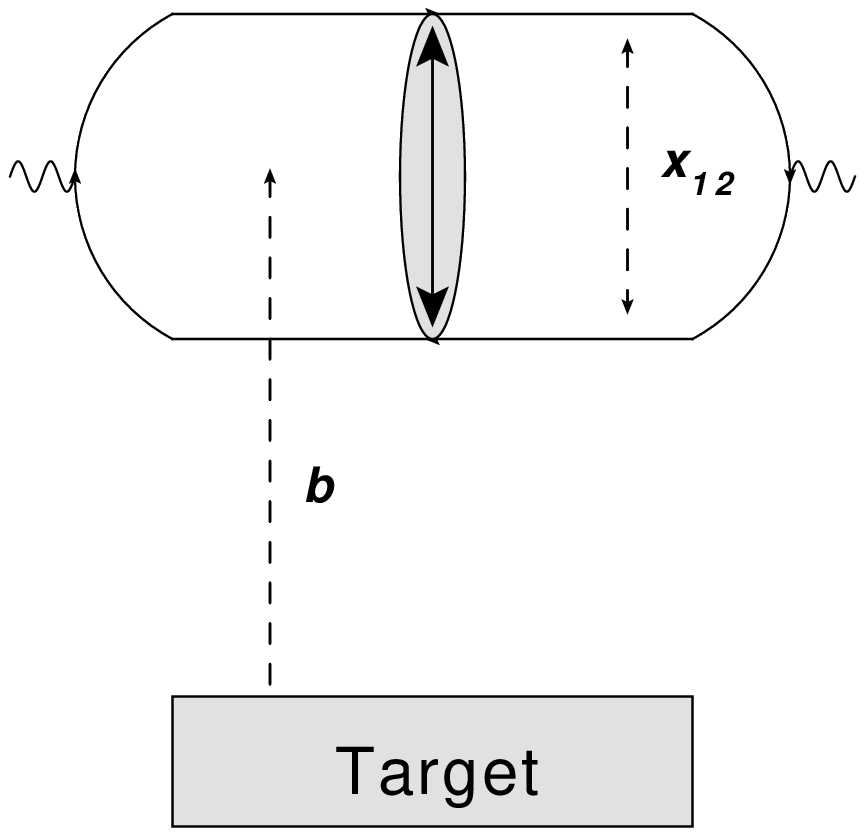}  & 
\hspace{2cm} &
\includegraphics[width=1.5in]{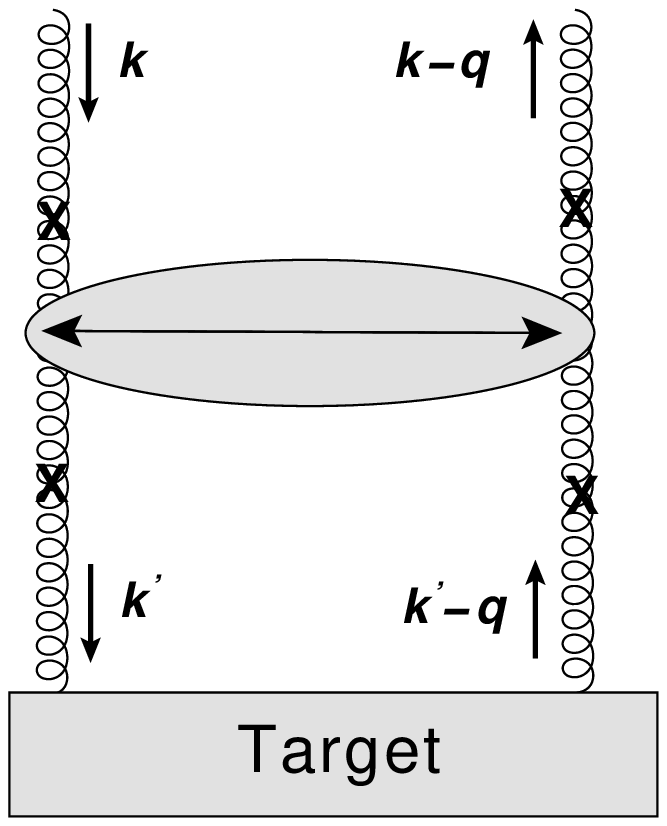}  \\
a) & & b)
\end{array}$
\end{center}
\caption{The substitution $k \leftrightarrow x_{12}$ rotates the BFKL Kernel in the transverse space from $a)$ $s$-channel in the dipole model  to $b)$ $t$-channel in the reggeized gluon formulation.  The arrows in the dark  blob on the both of pictures reflect the direction in which the BFKL Kernel operates. } \label{fig:dipolevsgluon}
\end{figure}

\hspace{0.4cm} Before considering the non-forward case we want to stress that in proving the duality symmetry we had zero transferred momentum $q=0$ together with a fixed impact parameter, in contrast to a conventional wisdom that tells us to integrate over $b$ for $q=0$. The reason for that is the fact that the transferred momentum is not dual to the impact parameter, but rather to the dipole displacement as we show below.  Let us recall that the dipole model exploits $s$-channel unitarity, whereas the reggeized gluon formulation is based on $t$-channel unitarity. This suggests that the substitution $k \leftrightarrow x_{12}$ rotates the BFKL Kernel in the transverse space as shown in Fig.~\ref{fig:dipolevsgluon}

\begin{figure}[h]
\begin{center}
\includegraphics[width=2in]{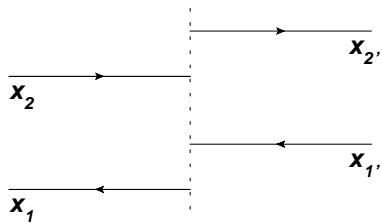}  
\end{center}
\caption{The initial dipole has different coordinates to the left and to the right to the unitarity cut (dashed line). The quark~(antiquark) line appears to be broken in the picture for a clearance of illustration, but there is no real discontinuity in the charge flow etc. The difference in coordinates happens in a natural way in the Fourier transform, if one keeps the conjugate momentum fixed. } \label{fig:nondiag}
\end{figure}

Rotating the Kernel (the dark blob) one can easily match the duals as follows.  The upper virtuality $k$ is dual to the dipole transverse size $x_{12}$ and  the impact parameter $b$ is dual to the lower virtuality $k'$.  To see the dual of the transferred momentum $q$ one has to imagine a dipole that has different sizes to the left $x_{12}$ and to the right $x_{1'2'}$ of the unitarity cut as shown in Fig.~\ref{fig:nondiag}.

Thus the transferred momentum $q$ is dual to the dipole displacement $x_{11'}$ (or $x_{22'}$ provided the size is not changed). 
The dipole model in its original formulation is dual to the forward case of the evolution of reggeized gluons. This is the reason why we fixed the impact parameter in  going from Eq.~(\ref{BFKLforward}) to Eq.~(\ref{BFKLdipole}). It is worth emphasizing that impact parameter $b$ as well as the lower virtuality $k'$ are related to the target, i.e. initial condition and thus are not relevant to the pure evolution that is general and holds for any initial condition. The lower virtuality $k'$ was introduced in the BFKL  equation to have target-projectile symmetry, whereas in the dipole model the evolution is build "from below" by construction and this fact is not obvious.

To see the duality symmetry in full, one has to consider an evolution equation for a dipole having different coordinates on the both sides of the unitarity cut, similar to that constructed in \cite{Levin:1900tt}. It was shown that the evolution of a non-diagonal dipole can be described by the evolution equation for some function $M(1,2:1,2')$ which has a meaning of a total cross section for  $x_{2}=x_{2'}$, namely,  
$M(1,2:1,2)=2N(1,2)$ (the optical theorem in the coordinate space). The resulting equation was solved, and its solution is found to be a linear combination of the BFKL solutions (the BK solutions for its non-linear version) $M(1,2:1,2')=N(1,2)+N(1,2')-N(2,2')$.  The meaning of this is that the non-diagonal dipole evolution can be described in terms of normal dipoles and does not require any introduction of a new colorless object (color quadrupole etc.). The complete analysis of the non-forward case will be published by us elsewhere.

Another important aspect of the duality symmetry $k \leftrightarrow x_{12}$ is that it relates in a very clear way the reggeization of gluons to the renormalization of the dipole wave function. The reggeization term in Eq.~(\ref{BFKLforward})
\begin{eqnarray}\label{reggezation}
 \int d^2\chi\frac{ k^2}{\chi^2(\chi-k)^2}\mathcal{F}(k) 
\end{eqnarray}
is, in fact, the term
\begin{eqnarray}\label{renorm}
 \int d^2 x_3\frac{ x_{12}^2}{x_{13}^2x_{23}^2}N(x_{12}) 
\end{eqnarray}
 of Eq.~(\ref{BFKLdipolezero}) which renormalizes the dipole wave function, due to the emission and absorption of a soft gluon before the interaction happens.

\section{Conclusion}
\label{sec:concl}

\hspace{0.5cm} It was shown  that the duality symmetry found by Lipatov \cite{Lipatov:1998as} solving the BFKL equation also holds for a non-linear case of the BK equation, provided we impose a constraint in the momentum space dual to the dipole size conservation. The duality symmetry can be viewed as a symmetry between the reggeized gluon formulation and the dipole picture of the high energy evolution. We give it a physical interpretation as a symmetry under rotation of the Kernel in the transverse space from $s$-channel~(the dipole model) to $t$-channel~(reggeized gluons).
The presented analysis shows that the dipole picture in its usual formulation for the vanishing impact parameter is dual to the forward case of the reggeized gluon formulation.

 We suggest the extension of the duality symmetry to the non-forward evolution and draw a physical picture for that case. In particular, we associate the initial dipole size $x_{12}$ with the upper virtuality of the BFKL ladder $k$, while the impact parameter $b$ is dual to the lower virtuality $k'$, and  the dipole displacement $x_{22'}$ is dual to the transferred momentum $q$.

The duality symmetry is related to the integrability of the system. The duality symmetry of the BK equation found in the present study can be regarded as an indirect indication of its integrability. A similar conclusion can be drawn from the crossing symmetry of the Kernel \cite{Korchemsky:1997fy}.

\subsection*{Acknowledgments}
\hspace{0.4cm}
We are deeply indebted to J.~Bartels, V.~Fadin, L.~Levin, L.~Lipatov and L.~Motyka for very helpful discussions. Special thanks go to G.~Korchemsky for his hospitality at Saclay where part of this work has been done. This study was supported by the Minerva Postdoctoral Fellowship of the Max Planck Society.

\section*{Appendix B} \label{appendix}

\hspace{0.4cm}We perform the Fourier transform of the non-linear term of the BK equation Eq.~(\ref{BK}) with impact parameter dependence suppressed.  
The calculation  is reduced to two terms, the squared dipole splitting  
\begin{eqnarray}\label{BKsquaredsplit}
 \int d^2 x_{3}  \frac{1}{x^2_{13}}N(x_{13})N(x_{32})
\end{eqnarray}

and the crossed splitting term
\begin{eqnarray}\label{BKcrossedsplit}
 \int d^2 x_{3}  \frac{x_{13}}{x^2_{13}}\frac{x_{23}}{x^2_{23}}N(x_{13})N(x_{32}).
\end{eqnarray}
Other two terms are obtained by  substitution $x_1 \leftrightarrow x_2$.
Let us consider first the squared splitting term of Eq.~(\ref{BKsquaredsplit}).
Using the conformal invariance of the BK amplitude we  pass to dimensionless coordinates given by  Eq.~(\ref{xcoord}) and Eq.~(\ref{kcoord})

\begin{eqnarray}\label{BKsquaredsplit2}
 \int d^2 \rho_{3}  \frac{1}{\rho^2_{13}}N(\rho_{13})N(\rho_{32}).
\end{eqnarray}

The Fourier transform defined in Eq.~(\ref{confFourier}) implies
\begin{eqnarray}\label{c1}
\int \frac{d^2 \chi_1 }{2\pi} e^{i \chi_1 \rho_{13}} \frac{d^2 \chi_2 }{2\pi}  e^{i \chi_2 \rho_{13}}\frac{\chi_1}{\chi^2_1}\frac{\chi_2}{\chi^2_2}
d^2 \chi_3 e^{i \chi_3 \rho_{13}}\mathcal{N}(\chi_3)
d^2 \chi_4 e^{i \chi_4 \rho_{23}}\mathcal{N}(\chi_4) \frac{1}{(2 \pi)^2}d^2 \rho_{12} e^{-i \rho_{12} \kappa} d^2 \rho_3
\end{eqnarray}
The integrations over $\rho_3$ and $\rho_{12}$ return $(2\pi)^2\delta^{(2)}(\chi_1+\chi_2+\chi_3+\chi_4)$ 
and $(2\pi)^2\delta^{(2)}(\chi_4+\kappa)$, respectively.  Integrated  over $\chi_3$ and $\chi_4$ the expression in Eq.~(\ref{c1}) reveals
\begin{eqnarray}\label{c2}
 \int d^2 \chi_1 d^2 \chi_2 \frac{\chi_1}{\chi^2_1} \frac{\chi_2}{\chi^2_2}  \mathcal{N}(\chi_1+\chi_2-\kappa) \mathcal{N}(\kappa).
\end{eqnarray}

There is one additional integration in Eq.~(\ref{c2}) compared to Eq.~(\ref{BKsquaredsplit}). 
As it was already mentioned, the basic assumption of the dipole model is that the dipole size does not change during rescatterings off the target. This condition is written implicitly in the dipole splitting Kernel by demanding that $x_{13}-x_{23}=x_{12}$. An analogous constraint is to be imposed on the momentum coordinates $\chi_3-\chi_4=\kappa$ by virtue of duality symmetry, because $\chi_3$, $\chi_4$ and $\kappa$ are conjugate to $x_{13}$, $x_{23}$ and $x_{12}$, respectively. This way we remove extra integration and find that $\chi_3=0$. Thus we conclude that the Fourier transform of Eq.~(\ref{BKsquaredsplit}) vanishes under constraint dual to the dipole size conservation. \\

The crossed term is Fourier transformed in a similar way. We rewrite it in $\rho$ coordinates 

\begin{eqnarray}\label{BKsquaredsplit2}
 \int d^2 \rho_{13}  \frac{\rho_{13}}{\rho^2_{13}}\frac{\rho_{23}}{\rho^2_{23}}N(\rho_{13})N(\rho_{32})
\end{eqnarray}
 and apply Eq.~(\ref{confFourier})
\begin{eqnarray}
\int \frac{d^2 \chi_1 }{2\pi} e^{i \chi_1 \rho_{13}} \frac{d^2 \chi_2 }{2\pi}  e^{i \chi_2 \rho_{23}}\frac{\chi_1}{\chi^2_1}\frac{\chi_2}{\chi^2_2}
d^2 \chi_3 e^{i \chi_3 \rho_{13}}\mathcal{N}(\chi_3)
d^2 \chi_4 e^{i \chi_4 \rho_{23}}\mathcal{N}(\chi_4) \frac{1}{(2 \pi)^2}d^2 \rho_{12} e^{-i \rho_{12} \kappa} d^2 \rho_3.
\end{eqnarray}
The integrations over $\rho_3$ and $\rho_{12}$ return $(2\pi)^2\delta^{(2)}(\chi_1+\chi_2+\chi_3+\chi_4)$ 
and $(2\pi)^2\delta^{(2)}(\chi_1+\chi_3-\kappa)$, respectively. After integration over $\chi_3$ and $\chi_4$ we obtain
\begin{eqnarray}\label{c4}
 \int d^2 \chi_1 d^2 \chi_2 \frac{\chi_1}{\chi^2_1} \frac{\chi_2}{\chi^2_2}  \mathcal{N}(\chi_1-\kappa) \mathcal{N}(\chi_2+\kappa).
\end{eqnarray}
Similar to the previous case we impose the constraint $\chi_3-\chi_4=\kappa$ and end up with 
\begin{eqnarray}\label{c4}
 \int d^2 \chi_1  \frac{\chi_1}{\chi^2_1} \frac{\chi_1-\kappa}{(\chi_1-\kappa)^2_2}  \mathcal{N}(\chi_1-\kappa) \mathcal{N}(\chi_1).
\end{eqnarray}
It is easy to see that the form of Eq.~(\ref{c4}) is identical to the expression in  Eq.~(\ref{BKsquaredsplit2}) we have started with. 
Identifying $\chi_1\rightarrow \rho_{13}$, $\kappa\rightarrow \rho_{12}$ and $\mathcal{N}(\chi) \rightarrow N(\rho_{12})$ this reads 

\begin{eqnarray}\label{crossedFourierFin}
  \int d^2 \rho_{3}  \frac{\rho_{13}}{\rho^2_{13}}\frac{\rho_{23}}{\rho^2_{23}}N(\rho_{13})N(\rho_{32})=\int d^2 x_{3}  \frac{x_{13}}{x^2_{13}}\frac{x_{23}}{x^2_{23}}N(x_{13})N(x_{32}).
\end{eqnarray}
As it was already mentioned in the text the non-linear term is a special one, it is invariant under Fourier transform with the constraint $\chi_3-\chi_4=\kappa$ and thus is self-dual. For example the linear BFKL terms  are mixed by Fourier transform, but their sum remains invariant under Eq.~(\ref{duality}).

\end{document}